\documentclass[twocolumn,english,aps,pra,nofootinbib,superscriptaddress,showpacs,showkeys,longbibliography]{revtex4-1}
\usepackage[T1]{fontenc}
\usepackage[latin9]{inputenc}
\usepackage{color}
\usepackage{babel}
\usepackage{bbold}
\usepackage{amssymb}
\usepackage{amsmath}
\usepackage{siunitx}
\usepackage{graphicx}
\usepackage{subfigure}
\graphicspath{ {images/} }
\usepackage{bbm, dsfont}
\usepackage{braket}
\usepackage[colorlinks=true,
    linkcolor=red, 
    citecolor=blue, 
    urlcolor =blue]{hyperref}
\makeatletter
\usepackage{amsfonts}
\usepackage{babel}
\usepackage{braket}
\newcommand\minus{
  \setbox0=\hbox{-}
  \vcenter{
    \hrule width\wd0 height \the\fontdimen8\textfont3
  }%
}

\makeatother

\begin{document}

\title{Rapid counter-diabatic sweeps in lattice gauge adiabatic quantum computing}
\author{Andreas Hartmann}
\affiliation{Institute for Theoretical Physics, University of Innsbruck, A-6020 Innsbruck, Austria}
\email{andreas.hartmann@uibk.ac.at}
\author{Wolfgang Lechner}
\affiliation{Institute for Theoretical Physics, University of Innsbruck, A-6020 Innsbruck, Austria}
\affiliation{Institute for Quantum Optics and Quantum Information of the Austrian Academy of Sciences, A-6020 Innsbruck, Austria}
\email{w.lechner@uibk.ac.at}

\begin{abstract}
We present a coherent counter-diabatic quantum protocol to prepare ground states in the lattice gauge mapping of all-to-all Ising models (LHZ) with considerably enhanced final ground state fidelity compared to a quantum annealing protocol. We make use of a variational method to find approximate counter-diabatic Hamiltonians that has recently been introduced by Sels and Polkovnikov [Proc. Natl. Acad. Sci. \textbf{114}, 3909 (2017)]. The resulting additional terms in our protocol are time-dependent local on-site y-magnetic fields. These additional Hamiltonian terms do not increase the minimal energy gap, but instead compensate for the Berry curvature. A single free parameter is introduced which is optimized via classical updates. The protocol consists only of local and nearest-neighbor terms which makes it attractive for implementations in near term experiments. 
\end{abstract}
\pacs{}
\maketitle

\section{Introduction}
A fundamental limitation in adiabatic quantum computation (AQC) is posed by the adiabatic theorem \citep{Born, Kato, messiah1961quantum} which states that a physical system follows its instantaneous eigenstate if the rate of change of a time-dependent Hamiltonian is much smaller than the energy gap between its lowest eigenstates. This inevitably results in a speed limit for any algorithm based on AQC, such as solving combinatorial optimization problems by quantum annealing \cite{Annealing2, Annealing1, ExperimentAnnealing, Farhi1, AdiabaticQuantumAnnealing, Annealing4, Annealing5, JohnsonQA, Annealing3, DWave, AnnealingTroyer, Rydberg}. As the minimal energy gap scales with the system size, the question whether a possible speedup in quantum annealing exists is thus still open.  

With the aim to overcome these fundamental limitations from the adiabatic condition, so called shortcut-to-adiabaticity methods \citep{STA} have been recently introduced. Shortcuts to adiabaticity (STA) make use of quantum coherence during the protocol and allow one to prepare the ground state or at least states that are close to the ground state in finite time. A variety of methods to engineer such protocols including invariant-based inverse engineering \citep{STA, DynamicalInvariants, Inverse, Inverse2}, fast-forward techniques \citep{NakamuraFF, NakamuraFF2011, MasudaFF, Masuda2012, TorronteguiFF2012, PhysRevLett.113.063003, JarzyinskiFF}, transitionless counter-diabatic driving \citep{Demirplak1, Demirplak2, Demirplak3, Berry, CDJarzyinski} and optimal control theory \citep{Optimal1,Optimal2,Optimal3} have been developed and applied to various fields such as quantum heat engines \citep{HeatEngine1, HeatEngine2, HeatEngine3}, atomic physics \citep{cirac2012goals, STA, delCampo5, TORRONTEGUI2013117}, open quantum systems \citep{opensystem, opensystem2, opensystem3, opensystem4, opensystem5}, Ising spin models \citep{delCampo2, Damski2014, Takahashi, SpeedupQA2017, HatomoraIsing, hatomura2018shortcuts, Steering}, adiabatic quantum computation \citep{AQCCounter, AQCCounter2} as well as experiments with spins and ions \citep{Exp1, ExpSchaff, ExperimentNature, ExperimentZhang, ExpColdIons2, CounterIonsExp2016}. Furthermore, the connection between the cost of STA protocols and its speedup compared to annealing protocols has been examined recently \citep{Cost1, SpeedCost, Cost3}.

However, for general spin glass models that solve optimization problems, counter-diabatic Hamiltonians contain non-local k-body interactions that are difficult to implement in experiment. Recently, a variational method to find approximate counter-diabatic protocols for arbitrary Hamiltonians has been introduced by Sels and Polkovnikov \citep{Polkovnikov}. An open challenge is to make use of these local counter-diabatic protocols in many-body systems with non-local all-to-all connectivity.

In this work, we present a counter-diabatic Hamiltonian consisting of single on-site local magnetic fields ($\sigma^x$, $\sigma^y$ and $\sigma^z$) and 4-body interactions between neighboring qubits ($\sigma^z\sigma^z\sigma^z\sigma^z$) to solve all-to-all connected combinatorial optimization problems. The scheme is based on the recently introduced encoding of optimization problems in a lattice gauge model (LHZ) \cite{LHZ} where the optimization problem is fully determined by local magnetic fields and problem-independent interactions among nearest-neighbor qubits. We present an approximate counter-diabatic protocol in LHZ by employing additional local single-body magnetic fields ($\sigma^y$) and derive the analytic expression for the time-dependent protocol as a function of local properties of each qubit. For small systems, we demonstrate numerically the implementation of rapid sweeps with large ground state fidelity and small excess energies compared to an adiabatic protocol.\\
A free control parameter for further improvement of the efficiency in a hybrid quantum-classical iterative update is introduced. The efficiency varies as a smooth function of this parameter even for short sweep times which allows for a simple variational optimization of the parameter. 

We note that the improved ground state fidelity does not stem from an increase of the minimal energy gap between ground and first excited state and cannot be understood in incoherent quantum annealing or with path integral Monte Carlo methods \citep{Troyer1, Troyer2, Inack2018}. The additional counter-diabatic term rather compensates for the Berry curvature which causes transitions between eigenstates \citep{Berry, Steering, BerryCurvature}. The method does not require a priori knowledge of the system's eigenstates which makes it feasible for experiment. Thus, the results further encourage the current efforts to build next generation quantum annealing experiments in the fully coherent regime.  

\section{Quantum Counter-Diabatic Annealing}
Quantum annealing aims at solving optimization problems which can be translated into Ising spin glasses with logical spins that can take the values $\pm 1$ \citep{Annealing1, Annealing2}. Finding the minimum energy of the spin glass is thus equivalent to determining the solution of the optimization problem \citep{LucasAnnealing}. The problem is cast into the form
$H_p=\sum_{i=1}^{N} \sum_{j < i} \tilde{J}_{ij} \tilde{\sigma}_i^z \tilde{\sigma}_j^z + \sum_{i=1}^{N} \tilde{b}_i \tilde{\sigma}_i^z$ where $\tilde{\sigma}_i^z$ is the z-Pauli matrix for the i-th logical spin and $N$ the total number of logical spins. The strengths of the magnetic fields $\tilde{b}_i$ and interactions $\tilde{J}_{ij}$ between the two sites $i$ and $j$ fully parametrize the system.
Starting in the ground state of a trivial initial state, for example $H_i=\sum_{i=1}^N \tilde{h}_i \tilde{\sigma}_i^x$ with $h_i$ the strength of the transverse magnetic field, the ground state of the problem Hamiltonian $H_p$ and thus the solution of our optimization problem is obtained by sufficiently slowly transferring the trivial initial state into the ground state of the problem Hamiltonian via the protocol
\begin{equation}
H(t)=f(t)H_i+(1-f(t))H_p, \label{eq:eq1}
\end{equation}
where $f(t)$ is a smoothly varying parameter in time with $f(t=0)=1$ and $f(t=\tau)=0$ at the beginning and end of the sweep, respectively. If the running time $\tau$ is infinitely large, the quantum system remains in its initial eigenstate for all times during the sweep.

To overcome this limitation from the adiabatic theorem, the counter-diabatic expression as a means of fast transitionless driving has been introduced by Demirplak and Rice \citep{Demirplak1, Demirplak2, Demirplak3} and Berry \citep{Berry}.\\
The basic idea of counter-diabatic driving (CD) is to evolve the system as 
\begin{equation}
H_{\textrm{CD}}(t)=H(t)+\dot{\lambda}(t) \mathcal{A}_{\lambda}(t), \label{eq:eq2}
\end{equation}
where $H$ is the Hamiltonian of Eq.\eqref{eq:eq1}, $\mathcal{A}_{\lambda}$ the adiabatic gauge potential and $\dot{\lambda}$ a single-component free control parameter. From now on, we will omit the explicit time-dependence in the description of the method.

Let us first highlight the role of the adiabatic gauge potential $\mathcal{A}_{\lambda}$ and control parameter $\dot{\lambda}$, respectively. A state $\ket{\psi}$ evolves under a time-dependent Hamiltonian $H \equiv H(\lambda)$ as $i \hbar \partial_t \ket{\psi} = H(\lambda) \ket{\psi} $ and $\ket{\tilde{\psi}}=U^{\dagger}\ket{\psi}$ in a rotating frame with respect to the time-dependent unitary transformation $U^{\dagger}$. The Hamiltonian in the rotating frame has the form
\begin{equation}
\tilde{H}_m=\tilde{H}-\dot{\lambda}\tilde{\mathcal{A}}_{\lambda}, \label{eq:eq3}
\end{equation}
where $\tilde{H}=U^{\dagger} HU$ is the diagonalized (stationary) instantaneous Hamiltonian and $\tilde{\mathcal{A}}_{\lambda}$ the adiabatic gauge potential in the rotating frame. The Hamiltonian $\tilde{H}$ is diagonal; thus all diabatic transitions occur due to the adiabatic gauge potential in the second term. Applying Eq.\eqref{eq:eq2} to Eq.\eqref{eq:eq3}, $\tilde{H}_{\textrm{CD,m}}=\tilde{H}$ is stationary and transitions get suppressed in the rotating frame such that the system remains in its instantaneous ground state for all velocities $|\dot{\lambda}|$ of the sweep. For a vanishing velocity $|\dot{\lambda}| \to 0$, the counter-diabatic Hamiltonian $H_{\textrm{CD}}$ coincides with the original Hamiltonian $H$ as expected.\\
The exact adiabatic gauge potential $\mathcal{A}_{\lambda}$ with $\hbar=1$ satisfies the following equation:
\begin{equation}
[i \partial_{\lambda} H-[\mathcal{A}_{\lambda},H], H]=0. \label{eq:eq4}
\end{equation}
Solving this equation for $\mathcal{A}_{\lambda}$ for the case of an Ising spin glass results in high order k-body interactions which are not realistic in current experimental implementations. Recently, a variational method with the aim to apply counter-diabatic terms to arbitrary Hamiltonians has been introduced by Sels and Polkovnikov \cite{Polkovnikov}. Here, one approximates the exact adiabatic gauge potential with an appropriate ansatz $\mathcal{A}^*_{\lambda}$. Solving Eq.\eqref{eq:eq4} for the approximate adiabatic gauge potential $\mathcal{A}^*_{\lambda}$ is equivalent to minimizing the Hilbert-Schmidt norm of the Hermitian operator
\begin{equation}
G_{\lambda}(\mathcal{A}^*_{\lambda})=\partial_{\lambda} H_0+i[\mathcal{A}^*_{\lambda},H_0] \label{eq:eq5}
\end{equation}
with respect to $\mathcal{A}^*_{\lambda}$ where we seek for the minimum of the operator distance \mbox{$D^2(\mathcal{A}^*_{\lambda})=Tr[(G_{\lambda}(\mathcal{A}_{\lambda})-G_{\lambda}(\mathcal{A}^*_{\lambda}))^2]$} between the exact $G_{\lambda}(\mathcal{A}_{\lambda})$ and approximate term $G_{\lambda}(\mathcal{A}^*_{\lambda})$. In turn, minimizing the operator distance is equivalent to minimizing the action 
\begin{equation}
\mathcal{S}(\mathcal{A}^*_{\lambda})=Tr[G^2_{\lambda}(\mathcal{A}^*_{\lambda})], \label{eq:eq6}
\end{equation}
associated with the approximate adiabatic gauge potential $\mathcal{A}^*_{\lambda}$, that is
\begin{equation}
\dfrac{\delta \mathcal{S}(\mathcal{A}^*_{\lambda})}{\delta \mathcal{A}^*_{\lambda}}=0, \label{eq:eq7}
\end{equation}
where $\delta$ denotes the partial derivative (see Ref.\citep{Polkovnikov} and \citep{GeometryPolkovnikov} for more details).

\section{Counter-diabatic Driving in the LHZ architecture}
In the recently introduced LHZ model \citep{LHZ}, the physical qubits describe the relative configuration of each two logical spins taking values 1 for parallel (i.e. $\uparrow \uparrow$, $\downarrow \downarrow$) and 0 for antiparallel ($\uparrow \downarrow$, $\downarrow \uparrow$) alignment, respectively. The time-dependent Hamiltonian in LHZ can be written in the form of Eq.\eqref{eq:eq1} as
\begin{align}
H_{\textrm{LHZ}}(t)&=\sum_{k=1}^{N_p} h_k(t) \sigma_k^x+\sum_{k=1}^{N_p} J_k(t)  \sigma_k^z \nonumber \\
&-\sum_{l=1}^{N_c} C_l(t) \sigma_{l,n}^z\sigma_{l,w}^z\sigma_{l,s}^z\sigma_{l,e}^z, \label{eq:eq8}
\end{align}
where $\sigma_k^{x}$ and $\sigma_k^{z}$ are the x- and z-Pauli matrices for the k-th physical qubit and the strengths of all local fields $h_k$, $J_k$ and constraints $C_l$, respectively, depend on time. Here, \mbox{$H_i(t)=\sum_{\textrm{k}}^{N_p} h_k(t) \sigma_k^x$} is the driver term and \mbox{$H_p(t)=\sum_{k}^{N_p} J_k(t)  \sigma_k^z-\sum_{l=1}^{N_c} C_l(t) \sigma_{l,n}^z\sigma_{l,w}^z\sigma_{l,s}^z\sigma_{l,e}^z$} the problem Hamiltonian to be solved.
The first two sums in Eq.\eqref{eq:eq8} run over all $N_p=N(N+1)/2-2$ physical qubits where $N$ is the number of logical spins in the original model and $h_k$ as well as $\tilde{J}_{ij} \to J_k$ are the strengths of controllable local fields that act on physical qubits. In the third sum, $C_l$ are the strengths of 4-body constraints constructed by closed loops of logical spins emerging due to the increased number of degrees of freedom from $N$ logical to $N_p$ physical qubits. To account for this, $N_c=N_p-2N+3$ four-body constraints among nearest neighbors on a square lattice are introduced. This notation includes $N_a=N-2$ auxiliary physical qubits in the bottom row of the LHZ architecture to obtain 4-body constraints on the whole square lattice. The indices $(l,n)$, $(l,w)$, $(l,s)$ and $(l,e)$ denote the northern, western, southern and eastern physical qubit of the constraint $l$, respectively (more details in Ref.\citep{LHZ}).

The sweep function $\lambda(t)$ is chosen to vanish at time $t=0$ and reaches $\lambda_f$ at time $t=\tau$ with
\begin{equation}
\lambda(t)=\lambda_0+(\lambda_f-\lambda_0)\sin^2\left(\dfrac{\pi}{2}\sin^2\left(\dfrac{\pi t}{2 \tau}\right)\right). \label{eq:eq9}
\end{equation}
Here, $\tau$ is the sweep time and $\lambda_0$ and $\lambda_f$ the values for initial and final time, respectively. This function $\lambda(t)$ has vanishing first and second order derivatives at the beginning and end of the sweep to attain smoothness of the function, i.e. $\dot{\lambda}(t=0)=\ddot{\lambda}(t=0)=\dot{\lambda}(t=\tau)=\ddot{\lambda}(t=\tau)=0$ and where
\begin{equation}
\dot{\lambda}(t)=(\lambda_f-\lambda_0)\dfrac{\pi^2}{4\tau}\sin\left(\dfrac{\pi}{\tau}t\right)\sin\left(\pi \sin^2 \left(\dfrac{\pi}{2\tau}t \right)\right) \label{eq:eq10}
\end{equation}
is the first time derivative of the protocol $\lambda(t)$.\\
In Hamiltonian \eqref{eq:eq8}, the time-dependent protocols for the strengths of the local fields $h_k(t)$, $J_k(t)$ and constraints $C_l(t)$ are of the same form as Eq.\eqref{eq:eq9} with initial and final values $h_{k,0}=1$,  $h_{k,f}=0$, $J_{k,0}=0$,  \mbox{$J_{k,f}=J_k$}, $C_{k,0}=0$ and $C_{l,f}=C_l$. The functions are explicitly given in the Appendix.

As a local and experimentally feasible ansatz for the adiabatic gauge potential $\mathcal{A}_{\lambda}$ of the LHZ Hamiltonian \eqref{eq:eq8}, we choose 
\begin{equation}
\mathcal{A}_{\lambda}^*=\sum_{i=1}^{N_p} \alpha_i \sigma_i^y \label{eq:eq11}
\end{equation}
where $\alpha_i$ is a time-dependent function to be determined.
The additional local magnetic field ($\sigma^y$) is introduced for each physical qubit.  This ansatz is imaginary; thus it breaks instantaneous time-reversal symmetry and adds a new degree of freedom to the system.\\
The operator \eqref{eq:eq5} in LHZ reads
\begin{align}
&G(\mathcal{A}_{\lambda}^*)=\sum_{k=1}^{N_p}(\dot{h}_k-2 \alpha_k J_k)\sigma_k^x+(\dot{J}_k+2 \alpha_k h_k)\sigma_k^z  \nonumber \\ 
&-\sum_{l=1}^{N_c} \dot{C}_l \sigma_{l,n}^z\sigma_{l,w}^z\sigma_{l,s}^z\sigma_{l,e}^z\nonumber \\
&+2C_l( \alpha_{l,n} \sigma_{l,n}^x\sigma_{l,w}^z\sigma_{l,s}^z\sigma_{l,e}^z+\alpha_{l,w} \sigma_{l,n}^z\sigma_{l,w}^x\sigma_{l,s}^z\sigma_{l,e}^z  \nonumber \\
&+\alpha_{l,s} \sigma_{l,n}^z\sigma_{l,w}^z\sigma_{l,s}^x\sigma_{l,e}^z+\alpha_{l,e} \sigma_{l,n}^z\sigma_{l,w}^z\sigma_{l,s}^z\sigma_{l,e}^x), \label{eq:eq12}
\end{align}
where the dot stands for the time derivative.
We compute the Hilbert-Schmidt norm by building the square of the Hermitian operator \eqref{eq:eq12}, that is
\begin{align}
&\dfrac{Tr[G_{\lambda}^2(\mathcal{A}_{\lambda}^*)]}{2^{N_p}}=\sum_{k=1}^{N_p}(\dot{h}_k-2 \alpha_k J_k)^2 + (\dot{J}_k+2 \alpha_k h_k)^2 \nonumber \\
& +\sum_{l=1}^{N_c} (\dot{C}_l)^2+4 C_l^2 (\alpha_{l,n}^2+\alpha_{l,w}^2+\alpha_{l,s}^2+\alpha_{l,e}^2), \label{eq:eq13}
\end{align}
where $2^{N_p}$ is the dimension of the Hilbert space.\\
The goal is to find an expression for $\alpha_k$ with minimal action in Eq.\eqref{eq:eq6} corresponding to a minimum in operator distance $D^2(\mathcal{A}^*_{\lambda})$ between exact and approximate adiabatic gauge potential. The optimal approximate solution for the adiabatic gauge potential $\mathcal{A}_{\lambda}^*$ is found by computing the derivative of the action with respect to $\alpha_{k}$ and applying Eq.\eqref{eq:eq7}. For the optimal solution we obtain
\begin{equation}
\alpha_k=\dfrac{1}{2}\dfrac{\dot{h}_k J_k-\dot{J}_k h_k}{J_k^2+h_k^2+\sum_n C_{k,n}^2}, \label{eq:eq14}
\end{equation}
where the sum in the denominator runs over all nearest neighbor constraints $C_{k,n}$ of the k-th physical qubit.\\
Note that this solution for the adiabatic gauge potential is exact for any constraint strength $C_l$ equal to zero, as it is just the counter-diabatic solution for $N_p$ independent two-level systems \citep{HatomoraIsing,Takahashi}. The adiabtic gauge potential $\mathcal{A}_{\lambda}^*$ also vanishes, if either $h_k=0$ or $J_k=0$ for all physical qubits, implying that the leading contribution to $\mathcal{A}_{\lambda}$ actually comes from the 4-body interaction terms.
For completeness, we can include 4-body interaction terms in our ansatz (see Appendix). The experimental implementation of the resulting 4-body terms is challenging and we will focus on the local solutions in this work.

The resulting local CD Hamiltonian in LHZ has the form
\begin{align}
&H_{\textrm{CD,LHZ}}(t)= \sum_{k=1}^{N_p} h_k(t) \sigma_k^x+\sum_{k=1}^{N_p} J_k(t)  \sigma_k^z \nonumber \\
&+\sum_{k=1}^{N_p} Y_k(\lambda_f, t) \sigma_k^y-\sum_{l=1}^{N_c} C_l(t) \sigma_{l,n}^z\sigma_{l,w}^z\sigma_{l,s}^z\sigma_{l,e}^z \label{eq:eq15}
\end{align}
where
\begin{align}
Y_k(\lambda_f, t) & = \alpha_k(t) \cdot \dot{\lambda}(\lambda_f, t) \nonumber \\
& = \dfrac{1}{2}\dfrac{\dot{h}_k(t) J_k(t)-\dot{J}_k(t) h_k(t)}{J_k^2(t)+h_k^2(t)+\sum_n C_{k,n}^2(t)} \cdot \dot{\lambda}(\lambda_f, t) \label{eq:eq16}
\end{align}
with $\dot{\lambda}(\lambda_f, t)$ as in Eq.\eqref{eq:eq10} and $\lambda_0 \equiv 0$. Note that in Eq.\eqref{eq:eq16} $\dot{\lambda}$ is the only term that depends on $\lambda_f$.

Equations \eqref{eq:eq15} and \eqref{eq:eq16}, together with the variational optimization of the  parameter $\lambda_f$ in the term $\dot{\lambda}$ of Eq.\eqref{eq:eq16} are the main results of this work. The complete implementation of this method reads as follows:
\begin{enumerate}
\item \textbf{Initial State:} Prepare the ground state of the trivial driver Hamiltonian $H_i$ and set an appropriate sweep time $\tau$. 
\item \textbf{CD sweep:} The strengths of the local fields $h_k(t)$, $J_k(t)$ and constraints $C_l(t)$ are driven according to protocol \eqref{eq:eq9} with initial and final values $h_k(0)=1$, $h_k(\tau)=0$, $J_k(0)=0$, $J_k(\tau)=J_k$, $C_l(0)=0$ and $C_l(\tau)=C_l$ for $k,l$ in the set of all physical qubits and constraints, respectively. Implement the protocol of the magnetic field strength in y-direction as in Eq.\eqref{eq:eq16}. 
\item \textbf{$\lambda_f$ values:} The global factor $\dot{\lambda}(t)$ in front of the counter-diabatic term \eqref{eq:eq16}, that is $\lambda_f$ in Eq.\eqref{eq:eq10} for the magnetic field strength in $\sigma^y$ with $\lambda_0 \equiv 0$, is optimized as a variational parameter from iterating the sweep in step 2 to maximize the final ground state fidelity.
\end{enumerate}

\section{Results}
In the first part of this section, we present our method with an example of Hamiltonian \eqref{eq:eq15} and a single randomly chosen instance $J_k$. In the second part, we present the statistics sampled from an ensemble of 100 randomly chosen instances. 

\subsection{Single instance}
Now, let us first describe the results for the local counter-diabatic Hamiltonian \eqref{eq:eq15} with an example of $N=4$ logical and thus $N_p=8$ physical qubits in LHZ. 
The strengths of the additional fields are bounded to $\lambda_f \in [\SI{-10}{J}, \SI{10}{J}]$.\\
As a measure for the efficiency we consider the squared instantaneous ground state fidelity \mbox{$F^2(t)=|\langle \psi(t)|\phi_0(t) \rangle|^2$} where $\phi_0(t)$ is the instantaneous ground state and $\psi(t)$ the state of the system at time $t$. 

\begin{figure}
\centering
\includegraphics[width=0.45\textwidth]{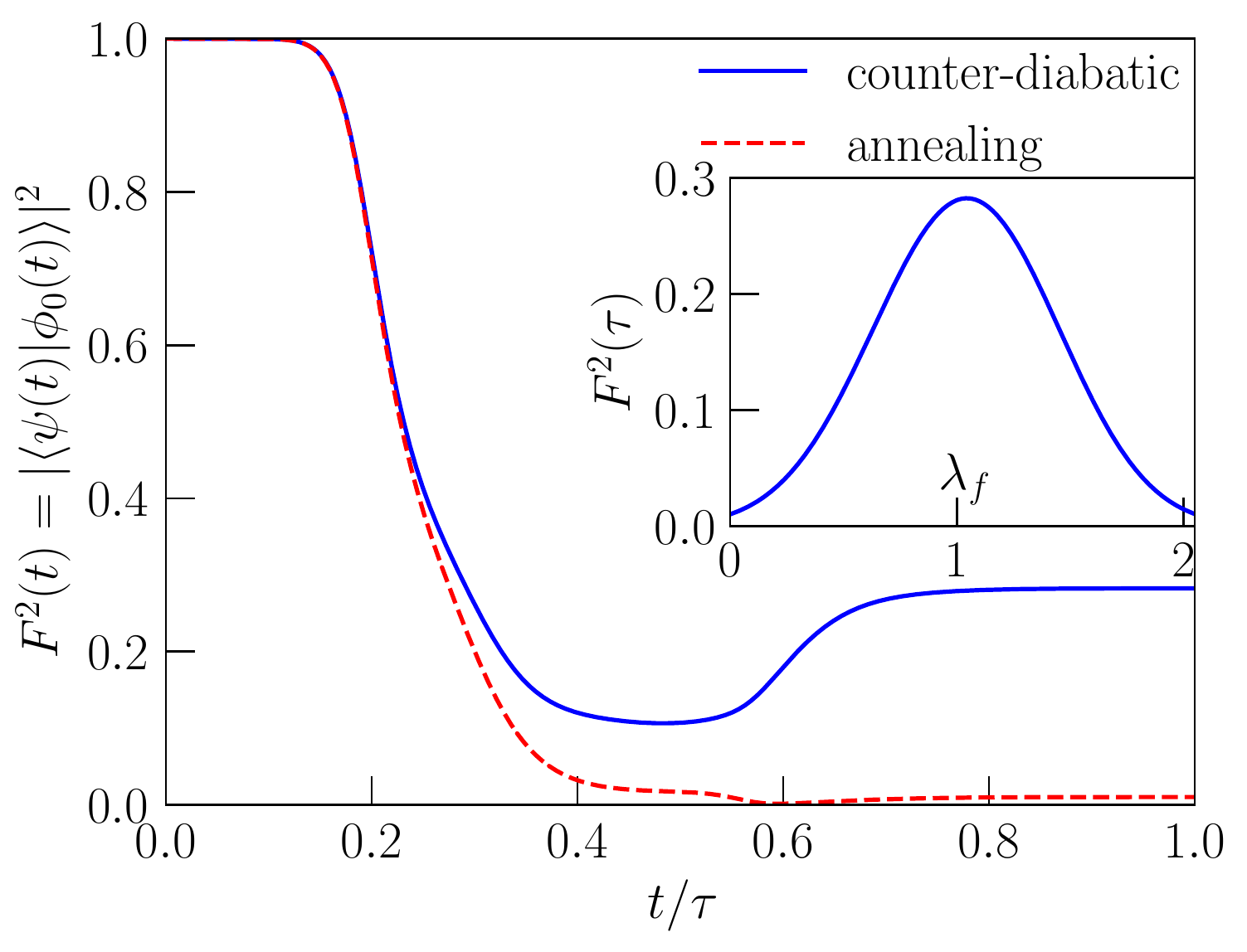}
\caption{\textbf{Counter-diabatic sweep.} The CD Hamiltonian \eqref{eq:eq15} in a LHZ architecture of $N_p=8$ physical qubits with parameters $h_k=\SI{1}{J}$ and $C_l=\SI{2}{J}$ for all physical qubits and constraints, respectively, auxiliary local field strength \SI{10}{J} and randomly uniformly distributed interaction strengths $J_k$ undergoes a counter-diabatic sweep with $\lambda_f=\SI{1.04}{J}$ and sweep time $\tau=\SI{1}{\per J}$. The main plot shows the squared instantaneous ground state fidelity during a sweep for the counter-diabatic and naive annealing case, respectively. The blue full and red dashed line correspond to the counter-diabatic \eqref{eq:eq15} and naive annealing Hamiltonian \eqref{eq:eq8}, respectively. The inset plot shows the distribution of the squared final ground state fidelity $F^2(\tau)$ over a set of different values of $\lambda_f$.}
\label{fig:fig1}
\end{figure}

Figure \ref{fig:fig1} depicts the squared instantaneous ground state fidelity $F^2(t)$ of a single instance of Hamiltonian \eqref{eq:eq15} with randomly uniformly distributed values of interaction strengths $J_k$ between $-1$ and $+1$ during a whole counter-diabatic sweep with $\tau=\SI{1}{\per J}$. 
For intermediate times during the sweep, the squared instantaneous ground state fidelity drops rapidly and then increases to a value of around $t/\tau = 0.28$, whereas for the naive annealing Hamiltonian \eqref{eq:eq8} rapidly decreases and then stays at a value of around 0.01.\\
The inset in Figure \ref{fig:fig1} depicts the free control parameter $\lambda_f$ in protocol \eqref{eq:eq10} which can dramatically enhance the performance of the counter-diabatic Hamiltonian \eqref{eq:eq15} and shows a Gaussian distribution of the squared final ground state fidelity $F^2(\tau)$ for multiple values of $\lambda_f$ close to its optimal value of \SI{1.04}{J}. The full width at half maximum (FWHM) is \SI{0.96}{J} which makes the local CD Hamiltonian \eqref{eq:eq15} stable against perturbations and allows for an experimental implementation of the iterative variational update.\par
\begin{figure}
\centering
\includegraphics[width=0.45\textwidth]{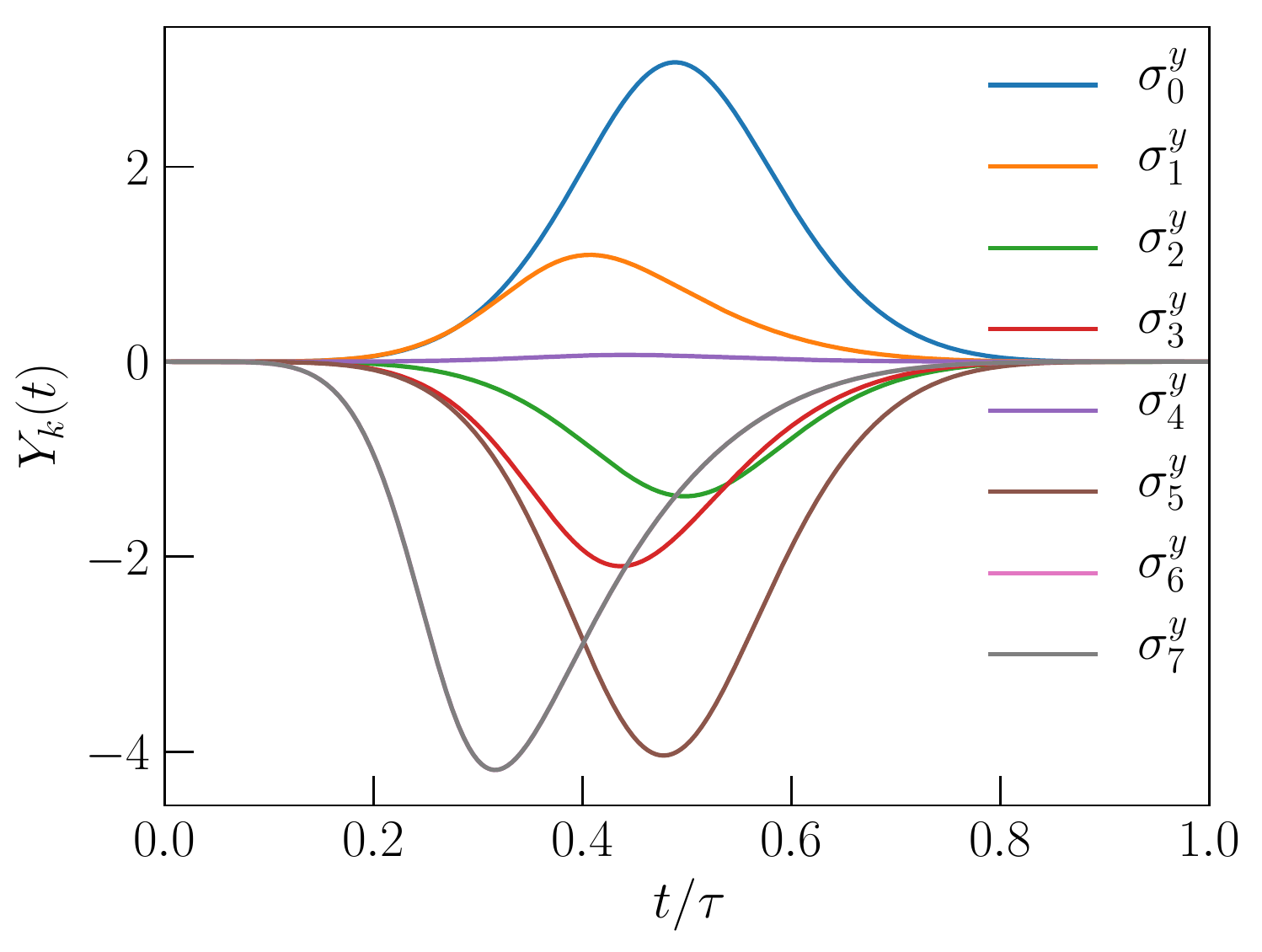}
\caption{\textbf{Regime of additional terms.} The strength of $Y_k(t)$ for each individual physical qubit in LHZ during a counter-diabatic sweep is shown.}
\label{fig:fig2}
\end{figure}

Figure \ref{fig:fig2} depicts the strengths of the additional magnetic fields in front of $\sigma_k^y$ for each physical qubit during this counter-diabatic sweep with the same parameters as described above.
The terms $\sigma_6^y$ and $\sigma_7^y$ correspond to the auxiliary physical qubits in the bottom row in the LHZ architecture. The strengths of the two auxiliary local magnetic fields are identical and their final values are each fixed to \SI{10}{J}.

\begin{figure}
\centering
\includegraphics[width=0.5\textwidth]{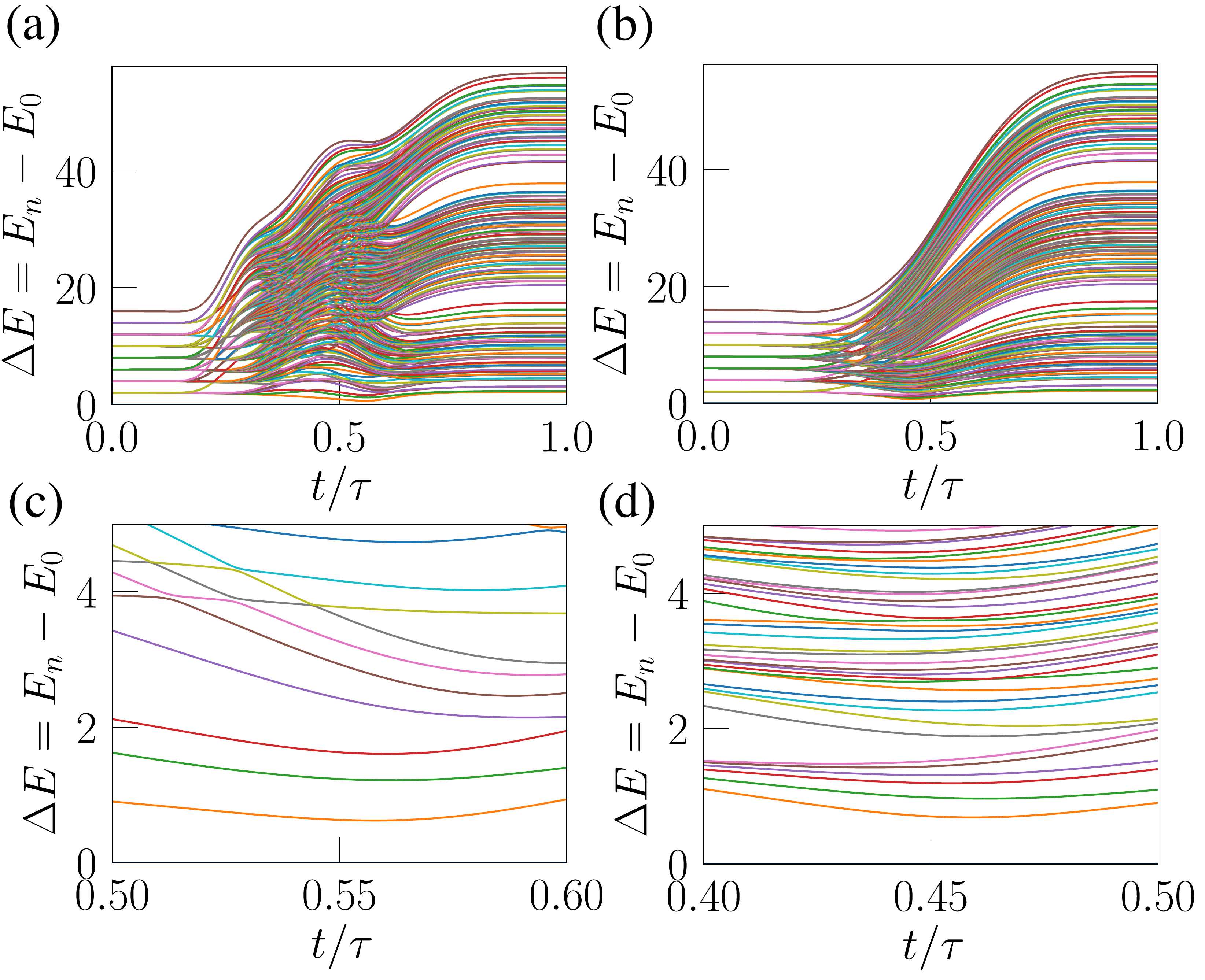}
\caption{\textbf{Minimal Energy Gap.} (a) and (c) show the energy spectrum of the counter-diabatic Hamiltonian \eqref{eq:eq15} with $\lambda_f=\SI{1.04}{J}$. (b) and (d) show the energy spectrum of the naive annealing Hamiltonian \eqref{eq:eq8}. The minimal energy gap $\Delta E_{min}$ between ground and first excited state shifts from around $t_{\textrm{A}} \approx 0.46 \tau$ for the naive annealing to $t_{\textrm{CD}} \approx 0.56 \tau$ for the counter-diabatic Hamiltonian. Note that even though the final ground state fidelity increases considerably, the minimal energy gap of the counter-diabatic Hamiltonian is smaller compared to that of the annealing Hamiltonian.}
\label{fig:fig3}
\end{figure}

In adiabatic protocols, the minimal energy gap is considered the fundamental limitation for the sweep time. We compare the energy spectra of the counter-diabatic Hamiltonian \eqref{eq:eq15} and annealing Hamiltonian \eqref{eq:eq8} with the same parameters as described above in Figure \ref{fig:fig3}. 
The position of the minimal energy gap $\Delta E_{min}=E_1-E_0$ between ground and first excited state shifts from around $t_{\textrm{A}} \approx 0.46 \tau$ for the naive annealing Hamiltonian \eqref{eq:eq8} to $t_{\textrm{CD}} \approx 0.56 \tau$ for the counter-diabatic Hamiltonian \eqref{eq:eq15} while the minimal energy gap even slightly decreased. This indicates that the minimal gap alone does not determine the efficiency in counter-diabatic protocols. 

\subsection{Statistical ensemble}
Let us now examine the statistics of counter-diabatic sweeps for an ensemble of randomly chosen instances.

\begin{figure}[htbp]
\centering
\includegraphics[width=0.5\textwidth]{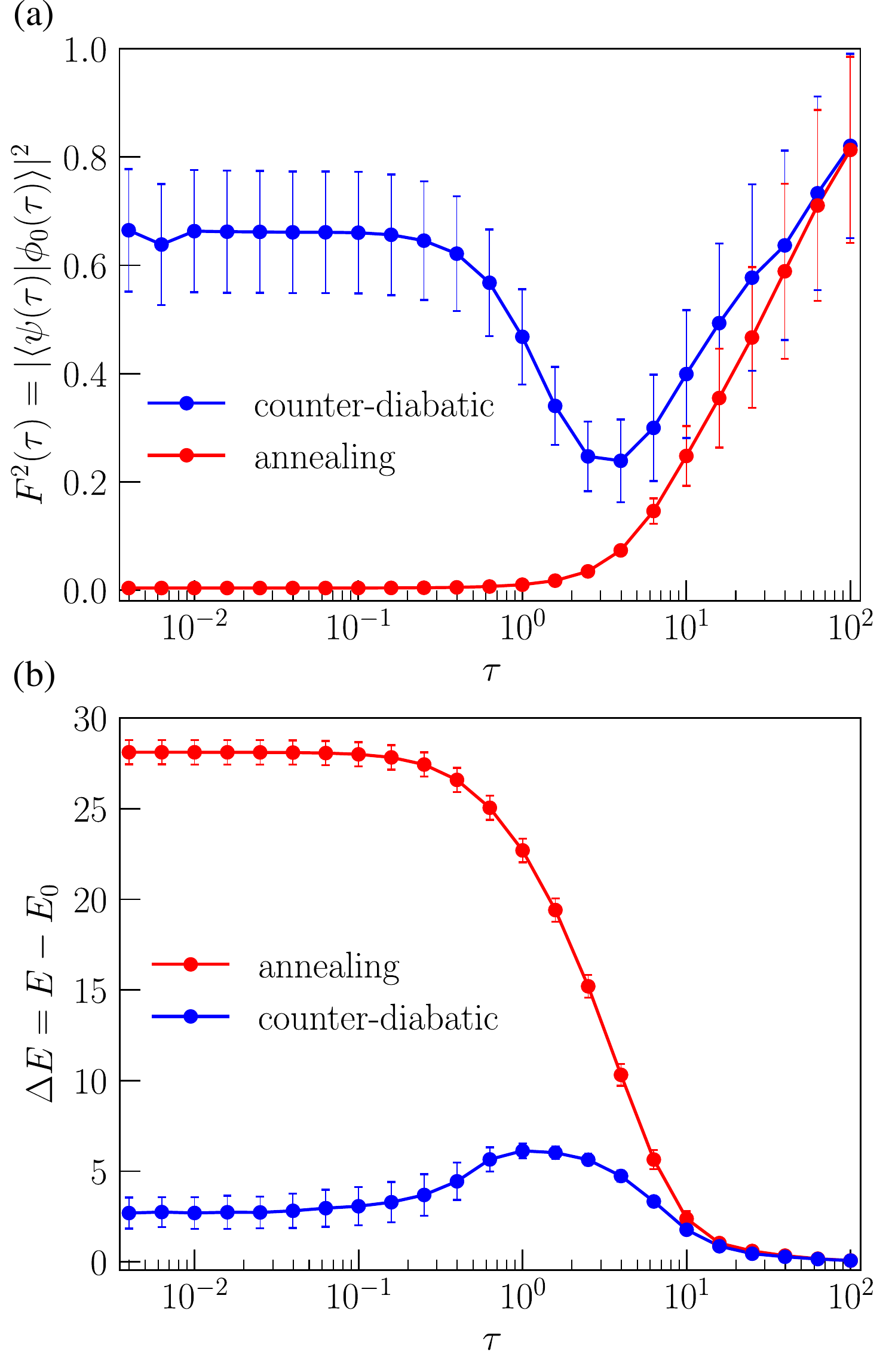}
\caption{\textbf{Ground state fidelities.} The CD Hamiltonian \eqref{eq:eq15} in LHZ with $N_p=8$ physical qubits and parameters $h_k=\SI{1}{J}$, $C_l=\SI{2}{J}$ and randomly uniformly chosen $J_k$ interaction matrices over 100 instances undergoes a counter-diabatic sweep. (a) shows the squared final ground state fidelity during different fast protocols $\tau$; (b) shows the excess energies during different fast protocols. The red circles are associated with the annealing protocol \eqref{eq:eq9} and the blue circles with the local counter-diabatic protocol \eqref{eq:eq16}. For both plots we have optimized the parameters $\lambda_f$ for each instance of $J_k$.}
\label{fig:fig4}
\end{figure}

Figure \ref{fig:fig4} depicts the mean squared final ground state fidelities $F^2(\tau)=|\langle \psi(\tau)|\phi_0(\tau) \rangle|^2$ and excess energies $\Delta E=E-E_0$, respectively, where averages are taken from protocols with fixed sweep times $\tau$ and uniformly distributed instances of $J_{k}$ in Hamiltonian \eqref{eq:eq15}.

In the quench limit $\tau \to 0$, the mean squared final ground state fidelity for the counter-diabatic Hamiltonian \eqref{eq:eq15} is around 0.67. For the annealing Hamiltonian \eqref{eq:eq8}, the probability of being in the final ground state is $1/2^8 \approx 0.0039$ which results in an enhancement of a factor of around 170. On the other hand, the corresponding excess energies of the counter-diabatic Hamiltonian is \SI{2.7}{J}; whereas for the naive annealing Hamiltonian they are approximately \SI{28}{J} which results in an improvement of a factor of around 10. For long running times, the final ground state fidelities and excess energies for both, counter-diabatic \eqref{eq:eq15} and naive annealing \eqref{eq:eq8} Hamiltonian, converge towards the same value as the amplitude of the added y-magnetic field becomes negligible compared to the annealing problem Hamiltonian for long running times due to the inversely proportional dependence of $\dot{\lambda}(t)$ on $\tau$. This behaviour also explains the drop for immediate long sweep times, as the relative improvement between counter-diabatic and annealing case decrease (see Figure \ref{fig:fig9} in the Appendix).

We note that there is an experimental limitation in implementing counter-diabatic protocols due to the cost of implementing the additional counter-diabatic term \eqref{eq:eq16}. This is a trade-off between the obtained increase in speed and energetic cost of our implemented CD Hamiltonian \eqref{eq:eq15} and its feasible applicability in the experiment.\\
Figure \ref{fig:fig5} depicts the energy scaling of the counter-diabatic term $Y_k(\lambda_f,t)$ \eqref{eq:eq16} and the derivative of the executed protocol \eqref{eq:eq10} for different fast protocols $\tau$. For different sweep times $\tau$, the control parameter $\dot{\lambda}(t)$ scales with the factor $1/\tau$. Considering the counter-diabatic protocol \eqref{eq:eq16}, enhancing the performance of the counter-diabatic Hamiltonian \eqref{eq:eq15} by one order of magnitude corresponds to an increase in the strength of the $Y_k(\lambda_f,t)$ term by a factor of 100.

\section{Conclusion and Outlook}
We have introduced an approximate optimal counter-diabatic driving protocol for the LHZ lattice gauge model architecture from a variational principle. Using an experimentally accessible local ansatz for the adiabatic gauge potential $\mathcal{A}_{\lambda}$, we derived a counter-diabatic Hamiltonian that consists of local fields only. This enables driving of counter-diabatic Hamiltonians even for all to-all connected spin glass problems. The counter-diabatic term $Y_k({\lambda_f,t)}$ added in Eq.\eqref{eq:eq15} is a local magnetic field in y-direction which depends on some free tuning parameter $\lambda_f$. This enables a hybrid classical-quantum algorithm where $\lambda_f$ is updated from measurements after the quantum process. 

Furthermore, we demonstrated a large increase in final ground state fidelity and decrease in excess energy with the counter-diabatic Hamiltonian \eqref{eq:eq15} compared to the annealing Hamiltonian \eqref{eq:eq8}. The CD driving keeps the system close to its ground state and dramatically enhances the performance of quantum annealing protocols to solve optimization problems using the lattice gauge model for Ising spins (LHZ). The increase in final ground state fidelity and decrease in excess energy, respectively, do not emerge due to an increase in the minimal energy gap; thus does not just follow the adiabatic theorem and Landau-Zener's formula. Instead, the position of the minimal energy gap shifts and our additional counter-diabatic term compensates for the Berry curvature which in general causes transitions between eigenstates. 

Remarkably, the ratio of the final ground state fidelity of counter-diabatic to annealing protocols seems to increase with the number $N_p$ of physical qubits (see Appendix for the example with $N_p=4$ physical qubits for comparison). While for $N_p=4$ we see an improvement by one order of magnitude it increases to a relative improvement of two orders of magnitude for $N_p=8$. This is an encouraging result which we will study in detail if this trend continues for larger systems.
\begin{figure}[htbp]
\centering
\includegraphics[width=0.3\textwidth]{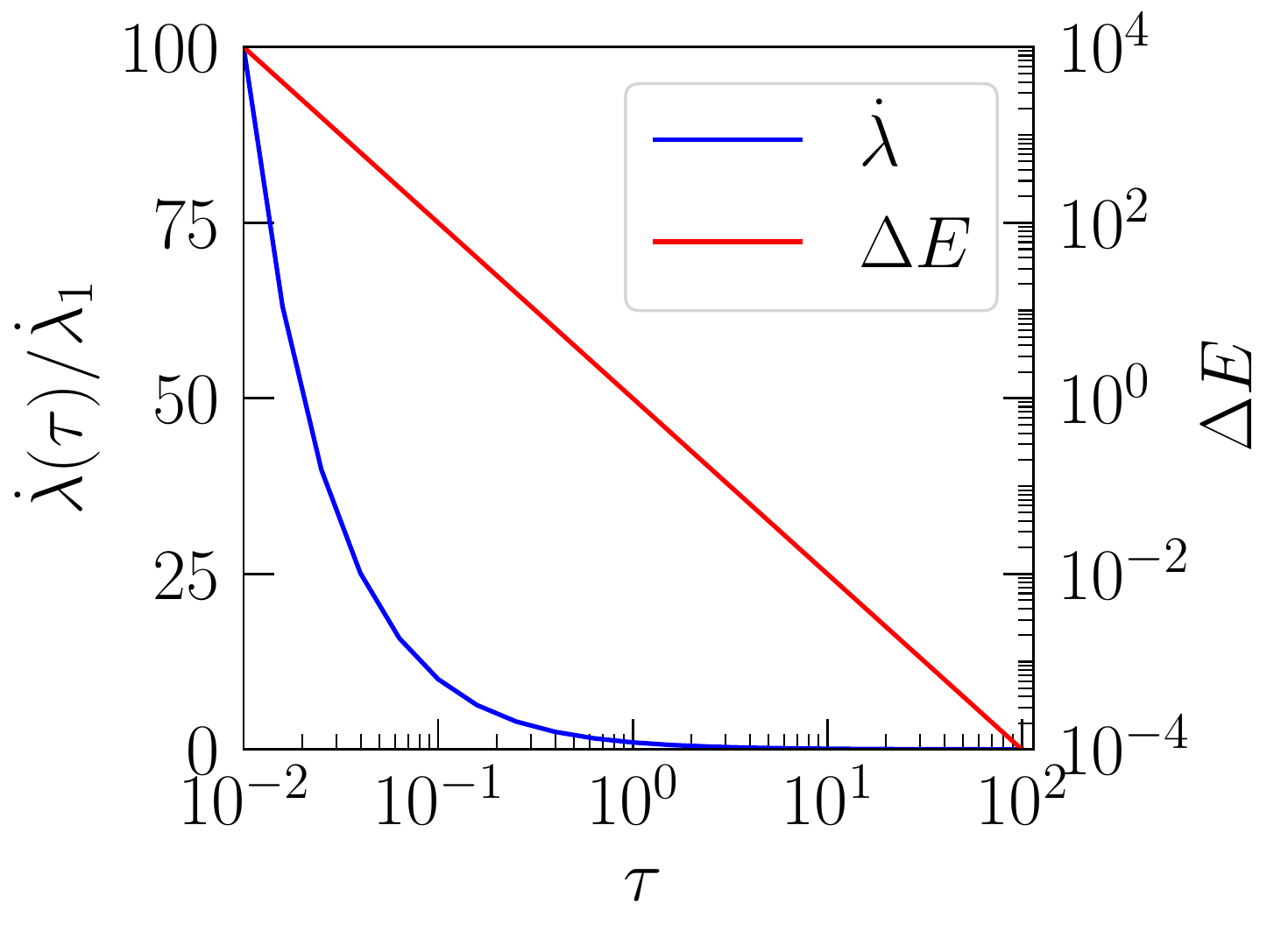}
\caption{\textbf{Energy scaling.} The energy scaling of the counter-diabatic term \eqref{eq:eq16} and protocol \eqref{eq:eq9} for different running times is shown.}
\label{fig:fig5}
\end{figure}

The energies of the $\sigma^y$ terms in Eq.\eqref{eq:eq16} scale with $1/\tau^2$ as shown in Figure \ref{fig:fig5}. Therefore, we expect an experimental limit for these counter-diabatic terms. The relevant and accessible regime is the one where the energy scale of the $\sigma^y$ term is comparable to that of the $\sigma^x$ term. In this regime, i.e. $\tau \approx \SI{e-1}{\per J}$, the resulting  speedup for solving optimization problems with counter-diabatic terms compared to annealing is around three orders of magnitude in the example above. 

We note that our method introduces a variational optimization of the parameter $\lambda_f$. In the regime of short running times where the $(\sigma^y)$ term dominates, the optimal value $\lambda_f^* \equiv \lambda_f/\tau$ is constant. For intermediate times, the parameter is optimized via variational optimization which has to be taken into account for the total time to solution. For the fastest switching times we considered, the energy scale of the counter-diabatic terms can be orders of magnitude larger compared to $J$ which may be the limiting factor for implementations. Improving the $\lambda_f$ optimization with respect to time to solution will be an interesting goal for future research. 

As a future direction, the counter-diabatic Hamiltonian \eqref{eq:eq15} of the LHZ architecture may be applied to quantum approximate optimization algorithms (QAOA) where the system is sequentially quenched with unitaries, that is, we can combine the speedup of the counter-diabatic protocol in the quench limit $\tau \to 0$ with the unitary quenches in QAOA \citep{farhi2014quantum, QAOAWolfgang} which may result in improved efficiency of this method.

The CD Hamiltonian \eqref{eq:eq15} is non-stoquastic \cite{bravyi2008complexity} and thus cannot be effectively solved with classical algorithms such as path integral Monte Carlo methods, but rather requires exact description of the full dynamics which also rules out simulations in stoquastic quantum annealing devices. We hope that our work contributes as a possible application to the current efforts in building next generation quantum annealing experiments with full quantum coherence. 

\section{Acknowledgements }
We thank Anatoli Polkovnikov, Dries Sels, and Kilian Ender for valuable discussions. The research was funded by the Austrian Science Fund (FWF) through a START grant under Project No. Y1067-N27 and the \mbox{Hauser-Raspe} foundation.

\newpage 

\newpage

\appendix*
\numberwithin{equation}{section} 
\section*{Appendix}
\setcounter{equation}{0}
\renewcommand{\theequation}{A.\arabic{equation}}

\subsection*{Derivation: Approximate adiabatic gauge potential}
Here we describe in detail the derivation of the adiabatic gauge potential in Eq.\eqref{eq:eq15} in the maintext.\\
Evolving a state $\ket{\psi}$ according to the Schr\"odinger equation $i \hbar \partial_t \ket{\psi} = H(\lambda) \ket{\psi} $ with a time-dependent Hamiltonian $H(\lambda) \equiv H$ in a rotating frame $\ket{\tilde{\psi}}=U^{\dagger} \ket{\psi}$ leads to
\begin{align}
&\tilde{H}_m \ket{\tilde{\psi}}= i \hbar \partial_t \ket{\tilde{\psi}} =i \hbar \partial_t(U^{\dagger} \ket{\psi}) \nonumber \\
&=i \hbar \partial_t U^{\dagger} \ket{\psi}+i \hbar U^{\dagger} \partial_t\ket{\psi} \nonumber \\
&=i \hbar \partial_t \lambda \, \partial_{\lambda} U^{\dagger} \ket{\psi}  + U^{\dagger} H \ket{\psi}  \nonumber \\
&=\partial_t \lambda \, (i \hbar \partial_{\lambda} U^{\dagger} U) \ket{\tilde{\psi}} + U^{\dagger} H U \ket{\tilde{\psi}} = (\tilde{H} - \dot{\lambda} \tilde{\mathcal{A}}_{\lambda}) \ket{\tilde{\psi}} \label{eq:eqA1}
\end{align}
where the adiabatic gauge potential reads
\begin{equation}
\mathcal{A}_{\lambda}=- i \hbar (\partial_{\lambda} U^{\dagger})U = i \hbar U^{\dagger} \partial_{\lambda}U. \label{eq:eqA2}
\end{equation}
Differentiating $\tilde{H}(\lambda)=U^{\dagger}(\lambda) H(\lambda) U(\lambda)$ with respect to $\lambda$, we obtain
\begin{align}
\partial_{\lambda}\tilde{H}=U^{\dagger} \partial_{\lambda} H U+\dfrac{i}{\hbar}[\tilde{\mathcal{A}}_{\lambda},\tilde{H}]. \label{eq:eqA3}
\end{align}
Going back to the laboratory frame, that is removing the tildes, and the fact that the gauge potential eliminates the off-diagonal terms of the moving Hamiltonian, i.e. $[\partial_{\lambda} \tilde{H}, \tilde{H}]=0$, we obtain
\begin{align}
[\partial_{\lambda} H + \dfrac{i}{\hbar} [\mathcal{A}_{\lambda},H],H]=0 \label{eq:eqA4}
\end{align}
where the first element in the commutator is precisely the operator $G(\mathcal{A}_{\lambda})$ of Eq.\eqref{eq:eq5} in the maintext with $\hbar=1$.\par
Using the ansatz $\mathcal{A}_{\lambda}^*=\sum_i \alpha_i \sigma_i^y$ of Eq.\eqref{eq:eq11} in the maintext and computing the operator $G(\mathcal{A}^*_{\lambda})$, leads to the commutator
\begin{align}
&i[A^*_{\lambda},H_{LHZ}]=\sum_{k=1}^{N_p} 2 \alpha_k h_k \sigma_k^z - 2 \alpha_k J_k \sigma_k^x \nonumber \\
&+\sum_{l=1}^{N_c} 2C_l (\alpha_{l,n} \sigma_{l,n}^x\sigma_{l,w}^z\sigma_{l,s}^z\sigma_{l,e}^z+\alpha_{l,w} \sigma_{l,n}^z\sigma_{l,w}^x\sigma_{l,s}^z\sigma_{l,e}^z \nonumber \\
&+\alpha_{l,s} \sigma_{l,n}^z\sigma_{l,w}^z\sigma_{l,s}^x\sigma_{l,e}^z+\alpha_{l,e} \sigma_{l,n}^z\sigma_{l,w}^z\sigma_{l,s}^z\sigma_{l,e}^x).
\label{eq:eqA5}
\end{align}
According to Eq.\eqref{eq:eq5} in the maintext, the derivative of $H_{\textrm{LHZ}}$ with respect to $\lambda$ (which in turn depends on time and thus just results in a time derivative) reads
\begin{align}
&\partial_{\lambda} H_{\textrm{LHZ}}(t)=\sum_{k=1}^{N_p} \dot{h}_k(t) \sigma_k^x+\sum_{k=1}^{N_p} \dot{J}_k(t)  \sigma_k^z \nonumber \\
&-\sum_{l=1}^{N_c} \dot{C}_l(t) \sigma_{(l,n)}^z\sigma_{(l,w)}^z\sigma_{(l,s)}^z\sigma_{(l,e)}^z.
\label{eq:eqA6}
\end{align}
Equations \eqref{eq:eqA5} and \eqref{eq:eqA6} combined give us our operator $G(\mathcal{A}_{\lambda}^*)$ of Eq.\eqref{eq:eq12} in the maintext.\par
For completeness, we can rotate the local, yet imaginary, CD Hamiltonian \eqref{eq:eq15} in the maintext in such a way that it becomes real. Applying the unitary rotation \mbox{$U(\theta)=exp \left(i \theta/2\sigma^z_k\right)=\cos\left(\theta/2\right)\mathbb{1}+i\sin\left(\theta/2\right)\sigma^z_k$} by a time-dependent angle $\theta$ to this imaginary Hamiltonian, that is
\begin{equation}
H_{\textrm{CD,rot}}=U H_{\textrm{CD}} U^{\dagger} + i (\partial_t U) U^{\dagger}, \label{eq:eqA7}
\end{equation} 
with the unitaries \mbox{$U^{\dagger}(\theta)=\cos\left(\theta/2\right)\mathbb{1}-i\sin\left(\theta/2\right)\sigma^z_k$}, \mbox{$\partial_t U(\theta)=-\dot{\theta}/2\sin\left(\theta/2\right)\mathbb{1}+i \dot{\theta}/2 \cos\left(\theta/2\right) \sigma^z_k$} and thus \mbox{$i (\partial_t U) U^{\dagger}=-\dot{\theta}/2 \sigma^z_k$}, the angle $\tan(\theta)=Y/X=Y_k/h_k$ and $\hbar=1$, we obtain the real counter-diabatic Hamiltonian
\begin{align}
&H_{\textrm{CD,real}}(t)=  \nonumber \\
&\sum_{k=1}^{N_p}\left(J_k(t)-\dfrac{1}{2}\dfrac{\dot{Y}_k(\lambda_f,t) h_k(t)-\dot{h}_k(t) Y_k(\lambda_f,t)}{h_k^2(t)+Y_k^2(\lambda_f,t)}\right)\sigma_k^z \nonumber \\
&+\sqrt{h_k^2(t)+Y_k^2(\lambda_f,t)}\;\sigma_k^x-\sum_{l=1}^{N_c} C_l(t) \sigma_{l,n}^z\sigma_{l,w}^z\sigma_{l,s}^z\sigma_{l,e}^z \label{eq:eqA8}
\end{align}
where we used the fact that $\dot{\theta}=d/dt\left(\arctan\left(Y_k/h_k\right)\right)=(\dot{Y}_k h_k-\dot{h}_k Y_k)/(h_k^2+Y_k^2)$ as well as \mbox{$\cos\theta=h_k/(\sqrt{h_k^2+Y_k^2})$} and $\sin\theta=Y_k/(\sqrt{h_k^2+Y_k^2})$.

\subsection*{Control parameter protocol}
\begin{figure}[htb]
\centering
\includegraphics[width=0.45\textwidth]{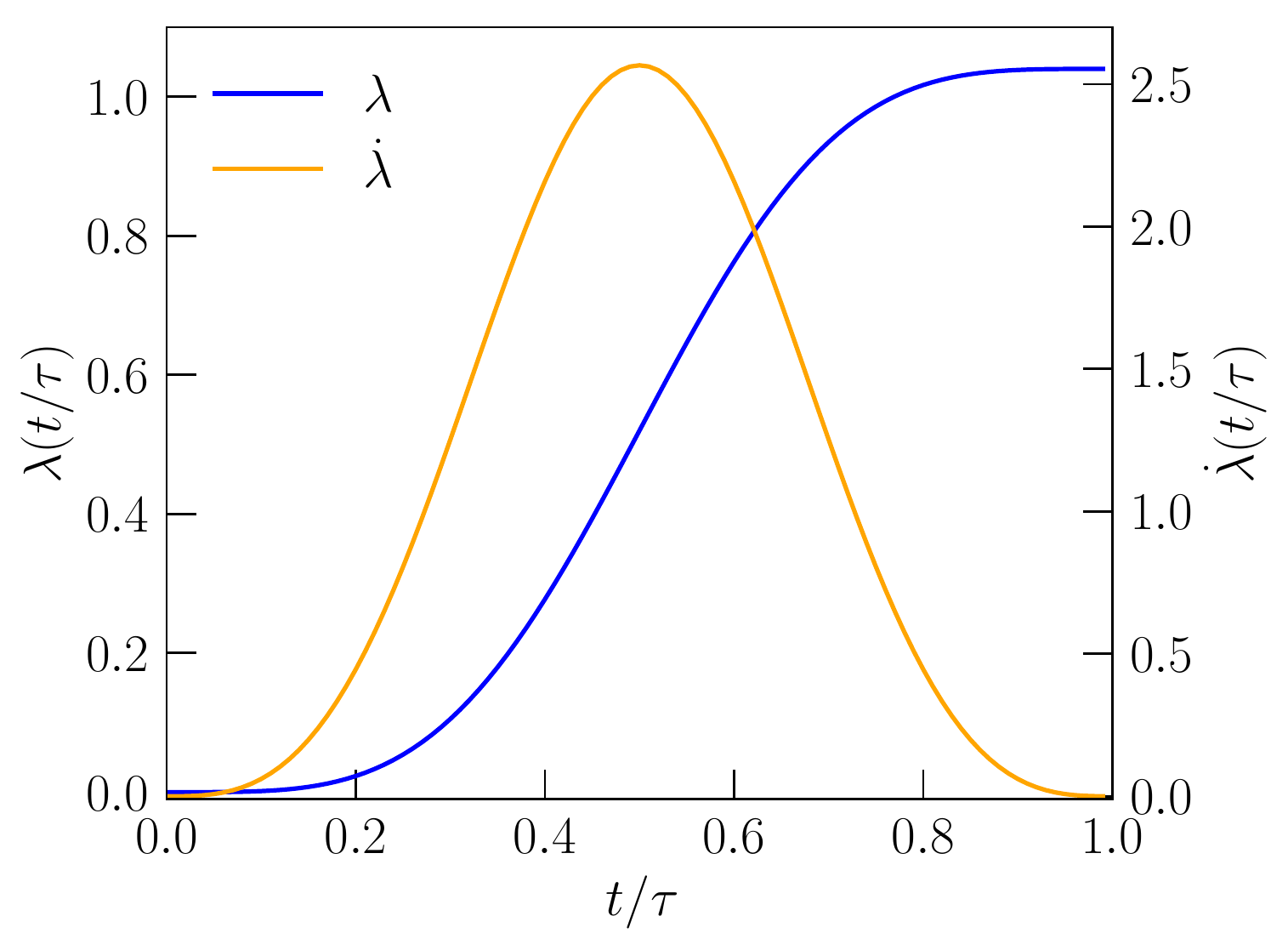}
\caption{\textbf{Control Parameter.} Protocols \eqref{eq:eq9} and \eqref{eq:eq10} are shown for initial and final values $\lambda_0=0$ and $\lambda_f=\SI{1.04}{J}$, respectively. $\dot{\lambda}(t/\tau)$ reaches its maximum value at $t=0.5 \tau$.}
\label{fig:fig6}
\end{figure}

Figure \ref{fig:fig6} depicts the normalized protocol $\lambda(t/\tau)$ and its derivative $\dot{\lambda}(t/\tau)$ for the case of $\lambda_0=0$ and \mbox{$\lambda_f=\SI{1.04}{J}$}, respectively.
Since the second derivative $\ddot{\lambda}(t/\tau)$ is zero at the beginning and end of the sweep, respectively, we attain smoothness of the function at these boundaries.

For the Hamiltonian \eqref{eq:eq8} in the maintext, the protocols for the strengths of the magnetic fields and constraints, respectively, have the explicit form
\begin{align}
h_k(t)&=h_{k,0}+(h_{k,f}-h_{k,0})\sin^2\left(\dfrac{\pi}{2}\sin^2\left(\dfrac{\pi t}{2 \tau}\right)\right) \nonumber \\
J_k(t)&=J_{k,0}+(J_{k,f}-J_{k,0})\sin^2\left(\dfrac{\pi}{2}\sin^2\left(\dfrac{\pi t}{2 \tau}\right)\right) \nonumber \\
C_l(t)&=C_{l,0}+(C_{l,f}-C_{l,0})\sin^2\left(\dfrac{\pi}{2}\sin^2\left(\dfrac{\pi t}{2 \tau}\right)\right)
\end{align}
and together with the time derivatives $\dot{h}_k(t)$ and $\dot{J}_k(t)$ are used in $\alpha_k$ (Eq.\eqref{eq:eq14} in the maintext).

\subsection*{Fidelity Distribution}
\begin{figure}
\centering
\includegraphics[width=0.45\textwidth]{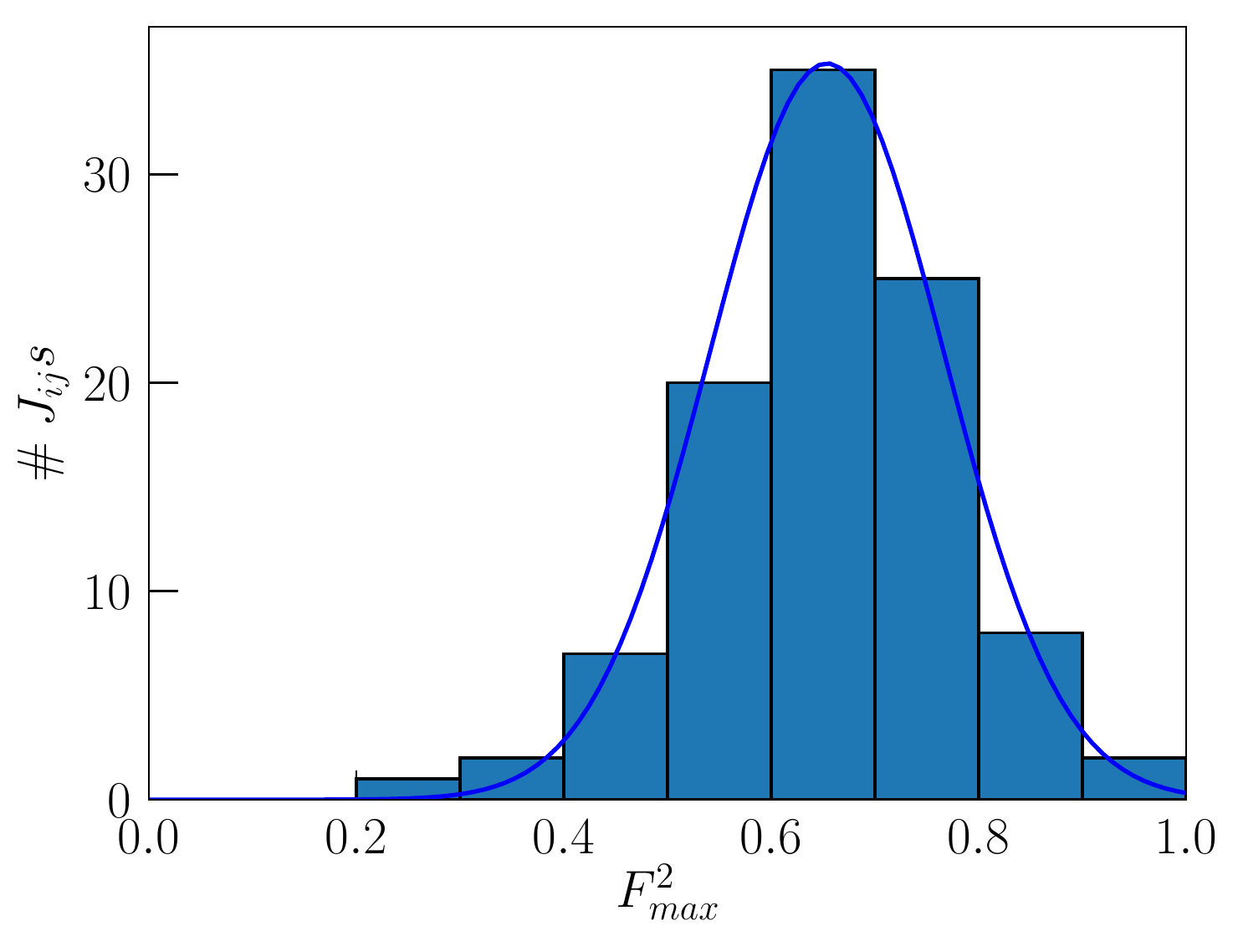}
\caption{\textbf{Distribution of the ground state fidelity.} An ensemble of 100 $J_k$ instances undergo a counter-diabatic sweep with sweep time $\tau=\SI{0.01}{\per J}$. The number of instances which reach a certain squared final ground state fidelity $F^2_{max}$ is shown. The blue line depicts a Gaussian function with mean $\mu=0.654$ and standard deviation $\sigma=0.113$.}
\label{fig:fig7}
\end{figure}

In the maintext, we have seen that for the quench limit $\tau \to 0$, the mean squared final ground state fidelity for the case of $N_p=8$ physical qubits is around 0.67.\\
Figure \ref{fig:fig7} shows the distribution of the reached maximal squared final ground state fidelities for all 100 $J_k$ instances for a very short running time $\tau=\SI{0.01}{\per J}$.
The distribution is roughly Gaussian with most of the instances having a final ground state fidelity between 0.60 and 0.70.

\subsection*{Minimal example: $N_p=4$}
\begin{figure}
\includegraphics[width=0.5\textwidth]{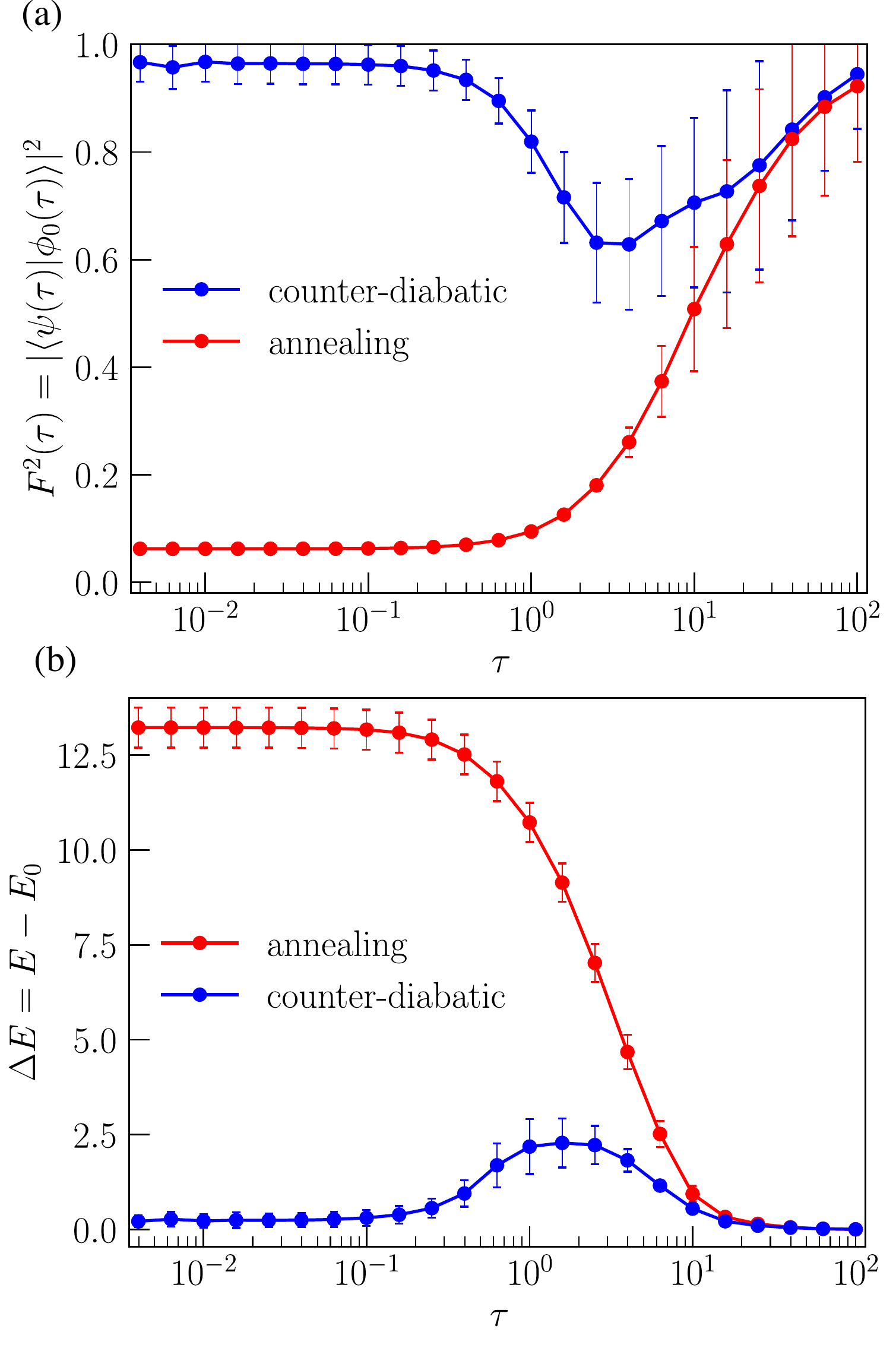}
\caption{\textbf{Ground state fidelities.} The CD Hamiltonian \eqref{eq:eq15} in LHZ with $N_p=4$ physical qubits and parameters \mbox{$C_l=\SI{2}{J}$}, $h_k=\SI{1}{J}$ and randomly chosen $J_k$ interaction strengths over 100 instances undergoes a counter-diabatic sweep. (a) shows the statistics of the squared final ground state fidelity during different fast protocols $\tau$; (b) shows the statistics of the excess energy during different fast protocols. The blue circles are associated with the counter-diabatic Hamiltonian \eqref{eq:eq15} and the red circles with the naive annealing Hamiltonian \eqref{eq:eq8}. For both plots we have optimized the bounded parameters $\lambda_f \in [\SI{-1000}{J},\SI{1000}{J}]$.}
\label{fig:fig8}
\end{figure}

\begin{figure}[htbp]
\centering
\includegraphics[width=0.45\textwidth]{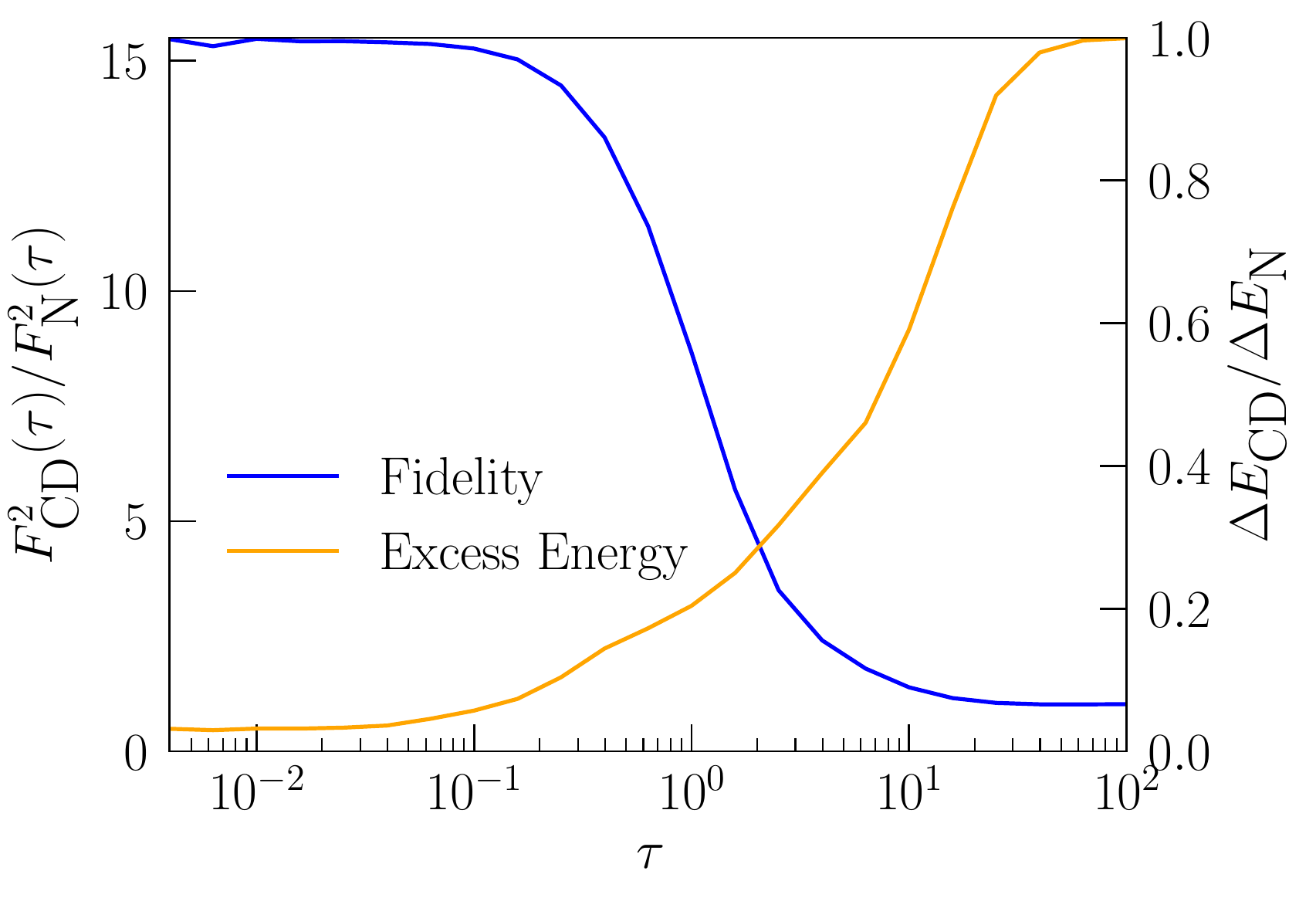}
\caption{\textbf{Relative improvement of the counter-diabatic protocol.} For $N_p=4$, an ensemble of 100 $J_k$ instances undergo counter-diabatic sweeps for different sweep times $\tau$. The ratio of the squared final ground state fidelity $F^2(\tau)$ of the counter-diabatic \eqref{eq:eq15} to naive annealing Hamiltonian \eqref{eq:eq8} is shown. The blue line depicts the ratio of the squared final ground state fidelities and the orange line the ratio of excess energies.}
\label{fig:fig9}
\end{figure}

With the aim to compare different system sizes, we consider the minimal example for LHZ with $N_p=4$ physical qubits. Figure \ref{fig:fig8} depicts the squared final ground state fidelities and excess energies for 100 randomly uniformly chosen $J_k$ instances for Hamiltonian \eqref{eq:eq15} in the maintext in a LHZ lattice gauge model with $N_p=4$ physical qubits and one constraint $C_1$ and where we have added one auxiliary qubit with local field strength \SI{10}{J} at the end of the sweep in the bottom row to obtain a 4-body constraint.
In the quench limit $\tau \to 0$, we achieve a squared final ground state fidelity of around 0.97 for the counter-diabatic Hamiltonian \eqref{eq:eq15} in the maintext and $1/2^4 = 0.0625$ for the annealing Hamiltonian \eqref{eq:eq8}, respectively, which gives an enhancement of a factor of around 15. The excess energy of the counter-diabatic Hamiltonian on the other hand is around \SI{0.4}{J}; whereas for the annealing case it stays at around \SI{13.2}{J} which gives an enhancement of a factor of around 33. Even in the quench limit, the probability to prepare the ground state of the CD Hamiltonian \eqref{eq:eq15} is finite.\\
The enhancement of the counter-diabatic protocol compared to annealing for different sweep times is measured via the squared final ground state fidelity as in the main text.
Figure \ref{fig:fig9} depicts the ratio between counter-diabatic and naive annealing squared final ground state fidelities and excess energies, respectively.
For the quench limit $\tau \to 0$, the ratio $F_{\textrm{CD}}(\tau)/ F_{\textrm{N}}(\tau)$ of counter-diabatic to naive annealing ground squared final ground state fidelities is around 15 and decreases to a value of 1. On the other hand, the ratio $\Delta E_{\textrm{CD}}/\Delta E_{\textrm{N}}$ of the excess energies of the counter-diabatic \eqref{eq:eq15} to the naive annealing Hamiltonian \eqref{eq:eq8} are around 0.03 in the quench limit and increase to a value of 1.

\subsection*{2nd order Ansatz}
As described in the maintext, if either $h_k=0$ or $J_k=0$ for all $N_p$ physical qubits, the adiabatic gauge potential $\mathcal{A}_{\lambda}^*$ vanishes and thus the leading contribution to the exact adiabatic gauge potential $\mathcal{A}_{\lambda}$ actually comes from the 4-body constraints which govern the dynamics of the system. Thus, we can include 4-body constraints with just odd numbers of imaginary pauli matrices $\sigma^y$ in our ansatz, that is,
\begin{align}
&\mathcal{A}_{\lambda}^*= \dfrac{1}{2} \sum_{i=1}^{N_p} \alpha_i \sigma_i^y +\sum_{l=1}^{N_c} \beta_l ( ^y_x\square^x_x + \,^x_x\square^y_x +\,^x_x\square^x_y+\,^x_y\square^x_x) \nonumber \\
&+\gamma_l (^y_z\square^z_z + \,^z_z\square^y_z +\,^z_z\square^z_y+\,^z_y\square^z_z) \nonumber \\
&+\delta_l ( ^y_x\square^y_y + \,^y_y\square^x_y +\,^y_x\square^y_y+\,^x_y\square^y_y) \nonumber\\
&+\epsilon_l (^y_z\square^y_y + \,^y_y\square^z_y +\,^y_z\square^y_y+\,^z_y\square^y_y). \label{eq:eqA9}
\end{align}
For abbreviation, the square $^1_4\square^2_3$ stands for the 4-body constraint $\sigma^1_{l,n}\sigma^2_{l,w}\sigma^3_{l,s}\sigma^4_{l,e}$ with $1,2,3,4 \in \{x,y,z\}$.\\
The action of the operator $G(\mathcal{A}^*_{\lambda})$ reads
\begin{widetext}
\begin{align}
&\dfrac{Tr[G^2(\mathcal{A}^*_{\lambda})]}{2^K}=\sum_{k=1}^{N_p} (\dot{h}_k-\alpha_k J_k)^2+(\dot{J}_k+\alpha_k h_k)^2 \nonumber \\
& + \sum_{l=1}^{N_c} ((h_{l,n}+h_{l,w}+h_{l,s}+h_{l,e}) \gamma_{l,n} -\dot{C}_{l,n})^2+(\alpha_{l,n} C_{l,n} -J_{l,n} \gamma_{l,n})^2 \nonumber \\
&+(\alpha_{l,w} C_{l,n} -J_{l,w} \gamma_{l,n})^2+(\alpha_{l,s} C_{l,n} -J_{l,s} \gamma_{l,n})^2+(\alpha_{l,e} C_{l,n} -J_{l,e} \gamma_{l,n})^2 \nonumber \\
&+\beta^2_{l,n} (J_{l,n}+J_{l,w}+J_{l,s}+J_{l,e})^2+\beta^2_{l,n} (h^2_{l,n}+h^2_{l,w}+h^2_{l,s}+h^2_{l,e}) \nonumber \\
&+(\delta_{l,n} (J_{l,n}+J_{l,w}+J_{l,s}+J_{l,e})-\epsilon_{l,n} (h_{l,n}+h_{l,w}+h_{l,s}+h_{l,e}))^2 \nonumber \\
&+4 \beta^2_{l,n} \sum_{m, 1 c.q.} C^2_{m,1 c.q.} +4 \delta^2_{l,n} \sum_{m} C^2_{m,1 c.q.} \nonumber \\
&+\gamma^2_{l,n} (|C_{m, c.q.}|) \sum_{m} C^2_{m}+4 \epsilon^2_{l,n} C^2_{l,n}+\epsilon^2_{l,n}(4-|C_{m, c.q.}|) \sum_{m} C^2_{m} \nonumber \\
&+(\beta_{l,n} (J_{l,n}+ J_{l,w})-\delta_{l,n} ( J_{l,s}+ J_{l,e}))^2+(\beta_{l,n} (J_{l,n}+ J_{l,s})-\delta_{l,n} ( J_{l,w}+ J_{l,e}))^2 \nonumber \\
&+(\beta_{l,n} (J_{l,n}+ J_{l,e})-\delta_{l,n} ( J_{l,w}+ J_{l,s}))^2+(\beta_{l,n} (J_{l,w}+ J_{l,s})-\delta_{l,n} ( J_{l,n}+ J_{l,e}))^2 \nonumber \\
&+(\beta_{l,n} (J_{l,w}+ J_{l,e})-\delta_{l,n} ( J_{l,s}+ J_{l,n}))^2+(\beta_{l,n} (J_{l,s}+ J_{l,e})-\delta_{l,n} ( J_{l,n}+ J_{l,w}))^2 \nonumber \\
&+(\epsilon_{l,n} (h_{l,n}+ h_{l,w})-\gamma_{l,n} ( h_{l,s}+ h_{l,e}))^2+(\epsilon_{l,n} (h_{l,n}+ h_{l,s})-\gamma_{l,n} ( h_{l,w}+ h_{l,e}))^2 \nonumber \\
&+(\epsilon_{l,n} (h_{l,n}+ h_{l,e})-\gamma_{l,n} ( h_{l,w}+ h_{l,s}))^2+(\epsilon_{l,n} (h_{l,w}+ h_{l,s})-\gamma_{l,n} ( h_{l,n}+ h_{l,e}))^2 \nonumber \\
&+(\epsilon_{l,n} (h_{l,w}+ h_{l,e})-\gamma_{l,n} ( h_{l,s}+ h_{l,n}))^2+(\epsilon_{l,n} (h_{l,s}+ h_{l,e})-\gamma_{l,n} ( h_{l,n}+h_{l,w}))^2 \nonumber \\
&+(\delta_{l,n} h_{l,n}-\epsilon_{l,n} J_{l,w})^2+(\delta_{l,n} h_{l,n}-\epsilon_{l,n} J_{l,s})^2+(\delta_{l,n} h_{l,n}-\epsilon_{l,n} J_{l,e})^2+(\delta_{l,n} h_{l,w}-\epsilon_{l,n} J_{l,n})^2 \nonumber \\
&+(\delta_{l,n} h_{l,w}-\epsilon_{l,n} J_{l,s})^2+(\delta_{l,n} h_{l,w}-\epsilon_{l,n} J_{l,e})^2+(\delta_{l,n} h_{l,s}-\epsilon_{l,n} J_{l,n})^2+(\delta_{l,n} h_{l,s}-\epsilon_{l,n} J_{l,w})^2 \nonumber \\
&+(\delta_{l,n} h_{l,s}-\epsilon_{l,n} J_{l,e})^2+(\delta_{l,n} h_{l,e}-\epsilon_{l,n} J_{l,n})^2+(\delta_{l,n} h_{l,e}-\epsilon_{l,n} J_{l,w})^2+(\delta_{l,n} h_{l,e}-\epsilon_{l,n} J_{l,s})^2
\label{eq:eqA10}
\end{align}
\end{widetext}
where $C_{m,1 c.q.}$ are the neighbor constraints of $C_l$ with 1 common qubit,  $|C_{m, c.q.}|$ is the number of common physical qubits of two constraints, that is either 1,2 or all 4 physical qubits share the same constraint, and $C_m$ the corresponding neighbor constraint (with a maximum of 8 nearest neighbor constraints plus the constraint itself).\\
Again like in first order, minimizing the action leads to the optimal solution. The derivative of $Tr[G^2(\mathcal{A}_{\lambda})]$ with respect to all parameters $\alpha_k,\beta_l, \gamma_l, \delta_l$ and $\epsilon_l$ and solving the linear equation system leads to the optimal, yet very unhandy, optimal solution and counter-diabatic Hamiltonian in 2nd order.

\end{document}